# Atomistic simulations of low-field mobility in Si nanowires: Influence of confinement and orientation

Neophytos Neophytou and Hans Kosina

Institute for Microelectronics, TU Wien, Gußhausstraße 27-29/E360, A-1040 Wien, Austria

e-mail: {neophytou|kosina}@iue.tuwien.ac.at

## Abstract

A simulation framework that couples atomistic electronic structures to Boltzmann transport formalism is developed and applied to calculate the transport characteristics of thin silicon nanowires (NWs) up to 12nm in diameter. The $sp^3d^5s^*$-spin-orbit-coupled atomistic tight-binding (TB) model is used for the electronic structure calculation. Linearized Boltzmann transport theory is applied, including carrier scattering by phonons, surface roughness (SRS), and impurities. We present a comprehensive investigation of the low-field mobility in silicon NWs considering: i) n- and p-type NWs, ii) [100], [110], and [111] transport orientations, and iii) diameters from $D$=12nm (electronically almost bulk-like) down to $D$=3nm (ultra-scaled). The simulation results display strong variations in the characteristics of the different NW types. For n-type NWs, phonon scattering and SRS become stronger as the diameter is reduced and drastically degrade the mobility by up to an order of magnitude depending on the orientation. For the [111] and [110] p-type NWs, on the other hand, large mobility enhancements (of the order of ~4X) can be achieved as the diameter scales down to $D$=3nm. This enhancement originates from the increase in the subband curvatures as the diameter is scaled. It overcompensates for the mobility reduction caused by SRS in narrow NWs and offers an advantage with diameter scaling. Our results may provide understanding of recent experimental measurements, as well as guidance in the design of NW channel devices with improved transport properties.

# I. Introduction

The recent advancements in process and manufacturing of nanoelectronic devices have allowed the manufacturability of nanowire (NW) devices, which are considered as candidates for a variety of applications. For field-effect-transistors, NWs have recently attracted large attention because of the possibility of enhanced electrostatic control, and transport close to the ballistic limit [1]. In that respect, ultra-scaled NW transistors of diameters down to $D$=3nm and gate lengths down to $L_G$=15nm have already been demonstrated by various experimental groups [2, 3, 4, 5, 6, 7]. Beyond ultra-scaled transistors, NWs have also attracted large attention as biological sensors [8], optoelectronic [9, 10], and very recently thermoelectric devices [11, 12].

Low-dimensional materials offer the capability of improved performance due to additional degrees of freedom in engineering their properties: i) the length scale of the cross section, ii) the transport orientation, and iii) the orientation of the confining surfaces [13, 14]. The carrier mobility that determines the performance of devices to a large degree, is a quantity closely related to the bandstructure via the effective masses, carrier velocities, and scattering rates. The carrier mobility in Si NWs as a function of the cross sectional size, transport orientation, and carrier type is the focus of this work.

The electronic properties of 1D silicon NWs [13, 14, 15, 16, 17, 18, 19] and 2D thin-body devices [20, 21, 22] in the high symmetry orientations [100], [110] and [111], have been addressed in a variety of theoretical studies. Effective mass models, with and without non-parabolic corrections [23], more sophisticated k•p models [21, 24], and atomistic tight-binding methods [13, 14, 15, 16, 17, 18, 19] have been utilized, both with unstrained and strained lattice. Ballistic and scattering dominated transport formalisms were also employed [23, 25, 26, 27]. These studies made clear that the challenges in simulating ultra-thin low dimensional devices in which the atoms in their cross section are countable, in general call for sophisticated models beyond the effective mass approximation. This is especially important for the valence band, where effective mass variations, valley splitting, and strong curvature variations are observed, which drastically affect the electronic properties [13, 28, 29].

Coupling sophisticated electronic structure models such as TB to proper transport models, however, has been proven a numerically difficult task, from both computation time and memory requirements. NW studies that couple TB to a variety of transport formalisms such as semiclassical ballistic [13, 20], quantum ballistic [16, 17], and scattering based transport (semiclassical or quantum) [19, 16, 30, 31, 32] are reported in the literature. Due to computational reasons, however, the structure size in such studies is limited to NWs of up to $D$=3nm in diameter.

In this work, we present a method to extract the low-field mobility of Si NWs by coupling the tight-binding (TB) model [13, 33, 34, 35] and linearized Boltzmann transport theory [36, 37] including phonon, surface roughness (SRS), and impurity scattering. The method is robust enough to capture atomistic bandstructure effects and computationally efficient to handle diameters up to $D$=12nm (~5500 atoms in the simulation domain), still on a single processor, or limited degree of parallelization. In this way, we can clearly observe the influence of the nanoscale effects on the electronic properties of the NWs as the diameter scales from $D$=12nm (electronically almost 3D bulk-like) down to $D$=3nm (1D like).

We then present a comprehensive study of the low-field mobility in NWs as a function of: i) diameter, ii) carrier type (n-type or p-type), and iii) transport orientation ([100], [110], [111]). Additionally, we explain how the transport distribution function, as described in [36], can provide intuitive understanding for the properties of NWs and serve as a convenient metric to compare the performance of different NW types. Such a comprehensive device exploration has not yet been reported. Our results could potentially provide explanations for the large mobility enhancement in certain NW categories that has been observed experimentally [38], the mobility trends in ultra-thin-layers [22], as well as useful insight into optimization strategies of NW channel transport devices. Specifically for p-type [110] and [111] NWs, confinement significantly decreases the effective mass of the subbands, and as a consequence the phonon-limited mobility increases by a factor of ~8X as the diameter scales from $D$=12nm down to $D$=3nm. This overcompensates the detrimental effect of enhanced SRS and phonon scattering from cross section scaling, factors which greatly degrade the mobility of n-type NWs.

This paper is organized as follows: In part II we describe the coupling of the atomistic TB method to the linearized Boltzmann transport approach. We derive the scattering rates for: i) phonons, ii) surface roughness, and iii) impurity scattering, and describe the approximations made to achieve computational efficiency. In part III we present the mobility analysis for n-type NWs and in part IV the analysis for p-type NWs. In part V we discuss the possible effect of surface relaxation on our results, and the cross section length scales at which 3D behavior appears, and finally, in part VI we summarize and conclude the work.

## II. Approach

### A. <u>Atomistic modeling:</u>

To obtain the bandstructure of the NWs both for electrons and holes for which spin-orbit coupling is important, a well calibrated atomistic model is used. The nearest neighbor $sp^3d^5s^*$-SO tight-binding model [13, 33, 34, 35] captures all the necessary band features, and in addition, is robust enough to computationally handle larger NW cross sections as compared to *ab-initio* methods. As an indication, the unit cells of the NWs considered in this study contain from ~150 to ~5500 atoms, and the computation time needed varies from a few to several hours for each case on a single CPU. Each atom in the NW unit cell is described by 20 orbitals including spin-orbit-coupling. The model itself and the parameterization used [33], have been extensively calibrated to various experimental data of various natures with excellent agreement [39, 40, 41, 42]. In particular, experimental data considering the valley splitting in slanted strained Si ultra-thin-body devices on disordered SiGe [41], and single impurities in Si [42], were matched without any material parameter adjustments. The model provides a simple but effective way for treatment of the surface truncation by hydrogen passivation of the dangling bonds on the surfaces [43]. What is important for this work, the Hamiltonian is built on the diamond lattice of silicon, and the effect of different orientations and cross sectional geometry is automatically included, all of which impact the interaction and mixing of various bulk bands. We consider cylindrical silicon NWs in three different transport orientations [100], [110], and [111] and diameters varying from $D$=12nm down to $D$=3nm. Relaxation of the NW surfaces is neglected in this study. In Part V we discuss the possible implications it might have to our results. The effect of strain engineering as a way to increase the carrier velocities and improve performance [44, 45] is not considered in this work.

The electronic structure of ultra-narrow NWs is sensitive to the diameter and orientation. For n-type NWs, we previously showed that the effective masses change as the diameter is scaled below 7nm, even up to 90% depending on the orientation [13]. More importantly, for p-type NWs we showed that the carrier velocities can change by a factor of ~2X with confinement (~4X variations in $m^*$), again depending on the orientation. In this case, the subband curvature is in addition sensitive to electrostatic potential variations in the cross section of the NWs, an effect that cannot be captured with simplified effective mass models [14, 15, 29]. This behavior is attributed to the strong anisotropy of the heavy-hole band. Indeed, the origin of most of these features can be traced back to and understood from the bandstructure of bulk Si. These dispersion variations will affect the mobility of the NWs.

### B. Boltzmann theory:

The Linearized Boltzmann transport theory is used to extract the electronic coefficients for each wire using its dispersion relation. We consider electron-phonon, surface roughness (SR), and impurity scattering. The electrical low-field conductivity ($\sigma$) follows from the linearized Boltzmann equation as:

$$\sigma = q_0^2 \int_{E_0}^{\infty} dE \left( -\frac{\partial f(E)}{\partial E} \right) \Xi(E), \tag{1}$$

where $\Xi(E)$ is the transport distribution function (TD) defined as [36, 37]:

$$\Xi(E) = \frac{1}{A} \sum_{k_x,n} v_n^2(k_x) \tau_n(k_x) \delta(E - E_n(k_x))$$
$$= \frac{1}{A} \sum_n v_n^2(E) \tau_n(E) g_{1D}^n(E). \qquad (2)$$

Here $v_n(E) = \frac{1}{\hbar} \frac{\partial E_n}{\partial k_x}$ is the bandstructure velocity, and $\tau_n(k_x)$ is the momentum relaxation time for a carrier with wavenumber $k_x$ in subband $n$,

$$g_{1D}^n(E_n) = \frac{1}{2\pi\hbar} \frac{1}{|v_n(E)|} \qquad (3)$$

is the density of states for the 1D subbands (per spin), and $A$ is the cross sectional area of the NW. The mobility ($\mu$) is defined as:

$$\mu = \frac{\sigma}{q_0 n}, \qquad (4)$$

where n is the carrier concentration in the channel. As we will show further on, the shape and magnitude of the transport distribution function is a very convenient way to provide intuitive explanations for the NW conductivity and mobility.

### C. Scattering rate calculation:

The transition rate $S_{n,m}(k_x, k_x')$ for a carrier in an initial state $k_x$ in subband $n$ to a final state $k_x'$ in subband $m$ is extracted from the atomistically extracted dispersions and waveform overlaps using Fermi's Golden Rule [46, 47] as:

$$S_{n,m}(k_x, k_x') = \frac{2\pi}{\hbar} |H_{k_x',k_x}^{m,n}|^2 \delta(E_m(k_x') - E_n(k_x) - \Delta E). \qquad (5)$$

The calculation of the relaxation rates involves an integral equation for $\tau_n(k_x)$ [21, 48, 49, 50]:

$$\frac{1}{\tau_n(k_x)} = \sum_{m,k_x'} S_{n,m}(k_x, k_x') \left(1 - \frac{v_m(k_x') \tau_m(k_x') f_m(k_x')}{v_n(k_x) \tau_n(k_x) f_n(k_x)}\right) \qquad (6)$$

While self-consistent solutions of this may be found, this is computationally very expensive, especially for atomistic calculations. Therefore, it is common practice to simplify the problem. For isotropic scattering (ADP, ODP, IVS) the term in the parenthesis reduces to unity upon integration over *k*-space due to symmetry considerations ($S_{n,m}$ is angle independent). In 1D the angle $\vartheta$ can take only two values: $\vartheta = 0$ and $\vartheta = \pi$. The momentum relaxation rates are, therefore, equal to the scattering rates. This holds for any bandstructure. For elastic intravalley, intrasubband scattering (even if anisotropic, i.e. SRS, or impurity scattering), and under the parabolic band assumption, the term simplifies to:

$$1 - \frac{v_m(k_x')\tau_m(k_x')f_m(k_x')}{v_n(k_x)\tau_n(k_x)f_n(k_x)} = 1 - \frac{v_m(k_x')}{v_n(k_x)} = 1 - \frac{\hbar v_m(k_x')/m_m^*}{\hbar v_n(k_x)/m_n^*} = 1 - \frac{|p_x'|}{|p_x|}\cos\vartheta \quad (7)$$

which is the usual term used in the calculation of the momentum relaxation times [47, 48]. Although this is strictly valid for intrasubband transitions only, it is often used for intersubband transitions as well (assuming a weak $k_x$–dependence of $\tau(k_x)$ [21]). In general, however, one should solve the full integral equation. As mentioned by Fischetti *et al.* in [21, 49], however, often sufficiently accurate results are obtained using the above approximations, without the need to evaluate numerically demanding integral equations. In this work, we calculate the relaxation times by:

$$\frac{1}{\tau_n(k_x)} = \sum_{m,k_x'} S_{n,m}(k_x,k_x')\left(1 - \frac{v_m(k_x')}{v_n(k_x)}\right) \quad (8)$$

The Fermi functions in Eq. (6) cancel for elastic processes. For inelastic, isotropic processes (as all the inelastic processes we consider) the term in brackets reduces to unity anyway after integration over $k_x$. Although admittedly in 1D $\tau(k_x)$ in the numerator and denominator of Eq. (6) can be varying with $k_x$, we still drop it as it is commonly done in the literature [19, 23, 30, 51]. This will only affect the SRS and impurity scattering results at larger diameters where intersubband transitions can be important, but at larger diameters the electronic structure approaches bulk-like, and the variation of $\tau(k_x)$ with $k_x$ is smaller.

The matrix element is computed using the scattering potential $U_S(\vec{r})$ as:

$$H_{k_x',k_x}^{m,n} = \int_{-\infty}^{\infty} \psi_{m,k_x'}^*(\vec{r})U_S(\vec{r})\psi_{n,k_x}(\vec{r})d^3r \quad (9)$$

where $\psi_{n,k_x}(\vec{r})$ is the wavefunction of a carrier in an initial state $k_x$ in subband $n$, and $\psi_{m,k_x'}(\vec{r})$ the wavefunction of a carrier in a final state $k_x'$ in subband $m$.

The total wavefunction of a specific state can be decomposed into a planewave in the $x$-direction, and a bound state in the transverse plane $\vec{R}$ as:

$$\psi_n(\vec{r}) = \frac{F_n(\vec{R})e^{ik_x x}}{\sqrt{\Omega}} \tag{10}$$

The matrix element then becomes:

$$H_{k_x',k_x}^{m,n} = \frac{1}{\Omega} \int_{-\infty}^{\infty} F_m^*(\vec{R}) e^{-ik_x' x} U_S(\vec{r}) F_n(\vec{R}) e^{ik_x x} d^2R\, dx. \tag{11}$$

### *1. Phonon scattering*:

In the case of phonon scattering, we extend the approach described in [47] for bulk and 2D carriers, to 1D carriers. The perturbing potential is defined as:

$$U_S(\vec{r}) = A_{\vec{q}} K_{\vec{q}} e^{\pm i(\vec{q}\cdot\vec{r} - \omega t)} \tag{12}$$

where $A_{\vec{q}}$ is associated with the lattice vibration amplitude and $K_{\vec{q}}$ with the deformation potential. The scalar product is split as $\vec{q}\cdot\vec{r} = \vec{Q}\cdot\vec{R} + q_x x$. In this case, the matrix element is:

$$H_{k_x',k_x}^{m,n} = \int_{-\infty}^{\infty} \frac{F_m^*(\vec{R})}{\sqrt{A}} \frac{e^{-ik_x' x}}{\sqrt{L_x}} A_{\vec{q}} K_{\vec{q}} e^{\pm i\vec{Q}\cdot\vec{R}} e^{\pm iq_x x} \frac{F_n(\vec{R})}{\sqrt{A}} \frac{e^{ik_x x}}{\sqrt{L_x}} d^2R\, dx, \tag{13}$$

where $L_x$ is the length of the unit cell. The integral over the transport direction $x$ becomes a Kronecker-delta expressing momentum conservation in the transport direction,

$$H_{k_x',k_x}^{m,n} = I_{k_x',k_x}^{m,n}(\vec{Q}) A_{\vec{q}} K_{\vec{q}} \delta_{k_x',k_x \pm q_x} \tag{14}$$

with

$$I_{k_x',k_x}^{m,n}(\vec{Q}) = \frac{1}{A} \int_R \rho_{k_x',k_x}^{m,n}(\vec{R}) e^{\pm i\vec{Q}\cdot\vec{R}} d^2R, \tag{15a}$$

$$\rho_{k_x',k_x}^{m,n}(\vec{R}) = F_{m,k_x'}(\vec{R})^* F_{n,k_x}(\vec{R}). \tag{15b}$$

When integrating and taking the square of the matrix element, the integral for the form factor is also evaluated as:

$$\left|I_{k_x',k_x}^{m,n}\right|^2 = \frac{1}{A^2}\int_R d^2R \int_{R'} d^2R' \, \rho_{k_x',k_x}^{m,n}(\vec{R})^* \rho_{k_x',k_x}^{m,n}(\vec{R}) e^{\pm i\vec{Q}\cdot(\vec{R}-\vec{R}')} \qquad (16)$$

The summation over the lateral momentum before substitution into Eq. (6) can be performed as:

$$\sum_{k_{\bar{R}}}\left|I_{k_x',k_x}^{m,n}\right|^2 = \frac{A}{4\pi^2}\frac{1}{A^2}\int_Q d^2Q\, e^{\pm i\vec{Q}\cdot(\vec{R}-\vec{R}')}\int_R d^2R \int_{R'} d^2R' \, \rho_{k_x',k_x}^{m,n}(\vec{R})^* \rho_{k_x',k_x}^{m,n}(\vec{R})$$

$$= A\int_R \left|\frac{\rho_{k_x',k_x}^{m,n}(\vec{R})}{A}\right|^2 d^2R \qquad (17)$$

$$= \frac{A}{A_{k_x',k_x}^{m,n}}$$

Here, $1/A_{k_x',k_x}^{m,n}$ has the units of m$^{-2}$. For computational efficiency of the calculation of the form factor overlap, on each atom we add the components of the squares of each multi-orbital wavefunction, and afterwards perform the final/initial state overlap multiplication. In such way, we approximate the form factor components by:

$$\sum_\alpha F_{n,k_x}^\alpha F_{m,k_x'}^{\alpha\,*} \sum_\alpha F_{n,k_x}^\alpha F_{m,k_x'}^{\alpha\,*} \approx \sum_\alpha F_{n,k_x}^\alpha F_{n,k_x}^{\alpha\,*} \sum_\alpha F_{m,k_x'}^\alpha F_{m,k_x'}^{\alpha\,*}, \qquad (18)$$

where $\alpha$ runs over the TB orbitals of a specific atom. This treatment is equivalent to computing the overlaps using the probability density of each state on each atomic side, as in a single orbital (i.e. effective mass) model, although we still keep the *k*-dependence of the wavefunctions. This makes,

$$\left|\rho_{k_x',k_x}^{m,n}(\vec{R})\right|^2 = \left|F_{m,k_x'}(\vec{R})^* F_{n,k_x}(\vec{R})\right|^2$$

$$\approx \sum_\alpha \left|F_{n,k_x}^\alpha(\vec{R})\right|^2 \sum_\alpha \left|F_{m,k_x'}^\alpha(\vec{R})\right|^2 \qquad (19)$$

$$\equiv \left\|F_{n,k_x}\right\|^2 \left\|F_{m,k_x'}\right\|^2$$

It reduces the memory needed in the computation by 20X, allowing simulations of large NW cross sections with only minimal expense in accuracy. In the cases of sine/cosine wavefunctions for square NWs, it can be shown analytically that for 1D intra-band the form factor becomes $\frac{1}{A_{nm}} = \frac{9}{4A}$ [27, 47]. For inter-band overlaps one obtains $\frac{1}{A_{nm}} = \frac{1}{A}$. Indeed, our numerical overlaps agree with these analytical expressions. To emphasize the importance of this approximation, we note that even after this simplification the storage of the probability density functions for the larger diameter NWs requires several Giga bytes of memory. For smaller diameters the overlap integrals calculated using the actual wavefunctions, or their probability distributions are in very good agreement. It would have been computationally prohibitive, however, to calculate them using the actual wavefunctions for the larger NWs, at least on a single CPU.

The transition rate is given by:

$$S_{n,m}(k_x, k_x') = \frac{2\pi}{\hbar} \left| I^{m,n}_{k_x',k_x} \right|^2 \left| K_{\vec{q}} \right|^2 \left| A_{\vec{q}} \right|^2 \delta_{k_x',k_x \pm q_x} \delta\left( E_m(k_x') - E_n(k_x) \pm \hbar\omega_{ph} \right) \quad (20)$$

where $\left| A_{\vec{q}} \right|^2 = \frac{1}{\Omega} \frac{\hbar \left( N_\omega + \frac{1}{2} \mp \frac{1}{2} \right)}{2\rho\omega_{ph}}$, $\rho$ is the mass density, and $N_\omega$ is the number of phonons given by the Bose-Einstein distribution. The relaxation rate of a carrier in a specific subband $n$ as a function of energy is then given by:

$$\frac{1}{\tau^n_{ph}(E)} = \frac{\pi}{\hbar} \frac{\left( N_\omega + \frac{1}{2} \mp \frac{1}{2} \right)}{\rho\hbar\omega_{ph}}$$

$$\times \left( \frac{1}{L_x} \sum_{m,k_x'} \frac{\left| K_{\vec{q}} \right|^2}{A^{k_x k_x'}_{nm}} \delta_{k_x',k_x \pm q_x} \delta\left( E_m(k_x') - E_n(k_x) \pm \hbar\omega_{ph} \right) \left( 1 - \frac{v_m(k_x')}{v_n(k_x)} \right) \right), \quad (21)$$

where $\hbar\omega_{ph}$ is the phonon energy, and we have used $\Omega = AL_x$. For acoustic deformation potential scattering (ADP or IVS), it holds $\left| K_{\vec{q}} \right|^2 = q^2 D_{ADP}^2$, whereas for optical deformation potential scattering (ODP for holes, IVS for electrons) $\left| K_{\vec{q}} \right|^2 = D_O^2$, where $D_{ADP}$ and $D_0$ are the scattering deformation potential amplitudes. For inter-valley scattering (IVS) in the conduction band we include all relevant $g$- and $f$-processes [47].

Specifically for elastic acoustic deformation potential scattering (ADP), after applying the equipartition approximation, the relaxation rate becomes:

$$\frac{1}{\tau_{ADP}^{n}(E)} = \frac{2\pi}{\hbar} \frac{D_{ADP}^{2} k_B T}{\rho v_s^2} \left( \frac{1}{L_x} \sum_{m,k_x'} \frac{1}{A_{nm}^{k_x k_x'}} \delta_{k_x',k_x \pm q_x} \delta\left(E_m(k_x') - E_n(k_x)\right) \left(1 - \frac{v_m(k_x')}{v_n(k_x)}\right) \right), \quad (22)$$

where $v_s$ is the sound velocity in Si.

The parameters we use are the same as in Ref. [47], with the exceptions of $D_{ODP}^{hole} = 13.24 \times 10^{10}\,\text{eV}/\text{m}$, $D_{ADP}^{hole} = 5.34\,\text{eV}$, and $D_{ADP}^{electron} = 9.5\,\text{eV}$ from Refs [19, 30, 46] which are more relevant for NWs. These are higher than the bulk values [47]. Larger deformation potential values are commonly used to explain mobility trends in nanostructures [23, 52, 53]. Various deformation potential parameter sets for nanostructures commonly appear in the literature [19, 52, 53]. Using different sets of parameters will change the absolute values of our results, but not the qualitative mobility trends with diameter and orientation, as we also show in Ref. [46] for p-type NWs. Buin *et* al. [19] calculated the mobility of NWs up to *D*=3nm using the same TB model and both confined and bulk phonons. In this work, by using the same parameters we were able to benchmark our results at least for the *D*=3nm bulk phonon case with a good agreement, before extending to larger diameters.

Confinement of phonons, and the details of their full dispersion are neglected. Dispersionless, bulk phonons, provide an ease of modeling and allow us to understand the effects of bandstructure on the mobility, still with good qualitative accuracy in the results. Spatial confinement mostly affects acoustic phonons, whereas optical phonons are not affected significantly by confinement [54, 55]. The degrading effect of confined phonons on the mobility for the thinnest NWs examined in this work can be of the order of 10-20%, and declines fast as the diameter increases [19, 30, 51, 54]. Especially for p-type thin layers, Donetti *et* al. have shown that using the full dispersion of confined phonons or using bulk phonons makes very little difference in mobility calculations [56]. Studies on GaAs NWs also reported similar conclusions [57]. These numbers, however, strongly depend on the boundary conditions one uses for the calculation of the phonon modes, which introduces an additional uncertainty [19, 54]. Clearly, the presence of additional scattering mechanisms (notably interface roughness scattering at such narrow diameters and especially impurity scattering at high impurity concentrations) also limits the mobility, and the relative effect of acoustic phonon confinement on the total mobility will be even smaller. Employing the full dispersion of acoustic confined phonons will only affect our results quantitatively. Using bulk phonons still captures the qualitative mobility trends with regards to geometry and orientation, which are mostly electronic structure related anyway. On the other hand, this quantitative reduction is partially captured by employing larger deformation potential parameters [23, 52, 53].

### *2. Surface roughness scattering*:

For SRS, we assume a 1D exponential autocorrelation function [58] for the roughness given by:

$$\langle \delta(\rho)\delta(\rho'-\rho) \rangle = \Delta_{rms}^2 e^{-\sqrt{2}|\rho'|/L_C} \qquad (23)$$

with $\Delta_{rms}$ = 0.48nm and $L_C$ = 1.3nm [23]. Surface roughness is assumed to cause a band edge shift. The scattering strength is derived from the shift in the band edges with quantization $\frac{\Delta E_{C,V}}{\Delta D}$ [59, 60], and the transition rate here is derived as:

$$S_{n,m}^{SRS}(k_x, k_x') = \frac{2\pi}{\hbar} \left( \frac{q_0 \Delta E_{C,V}}{\Delta D} \right)^2 \left( \frac{2\sqrt{2}\Delta_{rms}^2 L_C}{2 + q_x^2 L_C^2} \right) \delta\left( E_m(k_x') - E_n(k_x) \right), \qquad (24)$$

where $q_x = k_x - k_x'$.

This SRS model is more simplified than the ones described in Refs [23, 61, 62, 63] that account for the wavefunction deformation/shift at the interface, as well as additional Coulomb effects [64]. The band edge variation, however, is the dominant SRS mechanism in ultra-scaled channels as described by various authors, both theoretically and experimentally [23, 59, 60, 65, 66]. It has been shown in Ref. [23, 60] that the SRS limited low-field mobility in ultra-thin nanostructures follows a $D^6$ behavior, where $D$ is the confinement length scale, in this case the diameter of the channel. This is an indication that the dominant SRS contribution originates from the subband edge shift due to quantization. Jin *et* al. in Ref. [64] have considered additional Coulomb effects originating from the shift of the charge sheet near the interface, interface dipoles and image charges. They found that such effects can have significant contributions to SRS. In this work, however, we ignore these effects, and since they only cause quantitative changes in our results, we assume that they are lumped into an (enhanced) roughness $\Delta_{rms}$, which could be a reasonable approximation. Qualitative mobility trends in this work mostly originate from geometry-induced electronic structure variations. In addition, we investigate low-field mobility in NW channels without gate fields, potential variations in their cross section, or temperature variations. The probability density in such case is relatively flat near the interface of the NW and mostly concentrated in the middle of the channel, even for the smaller diameter NWs [14], which partially justifies our approximation.

The $D^6$ behavior originates from simple 1D quantization, where it follows, $\frac{\delta E_0}{\delta W} = \frac{\delta\left(\hbar^2\pi^2/2mW^2\right)}{\delta W} \sim \frac{1}{W^3}$, with $W$ being the width of the box. Therefore, the squared matrix element for the NWs follows a $D^{-6}$ dependence due to roughness. This differential change in the band edges $\frac{\Delta E_{C,V}}{\Delta D}$ versus diameter for NWs in different orientations is shown in Fig. 1a for the conduction bands (for both $\Gamma$ and off-$\Gamma$ valleys) and Fig. 1b for the valence bands. The dash-black lines indicate the $D^{-3}$ law. The band-edge variations increase as the diameter scales, following the $D^{-3}$ behavior. This is more evident for the conduction bands and the [100] valence band, whereas the band variation in the p-type [110] and [111] NWs is weaker with diameter reduction. For p-type, this orientation dependence originates from the highly anisotropic heavy-hole band, which determines the confinement effective masses.

### *3. Impurity scattering:*
Assuming the *x* direction to extend to infinity, the scattering potential for screened Coulomb ionized impurities is given by:

$$U_S(\vec{r}) = \frac{q_0^2}{4\pi\kappa_s\varepsilon_0} \frac{e^{-\sqrt{(\vec{R}-\vec{R}')^2+x^2}/L_D}}{\sqrt{(\vec{R}-\vec{R}')^2+x^2}}, \quad (25)$$

where $\vec{R}$ is the position of an electron in the 2D cross section at $x=0$, feeling the influence of an impurity at $(x, \vec{R}')$. The screening length $L_D$ is given by:

$$L_D = \sqrt{\frac{\kappa_s\varepsilon_0 k_B T}{q_0^2 \text{n}} \frac{\Im_{-1/2}(\eta_F)}{\Im_{-3/2}(\eta_F)}}. \quad (26)$$

where $\Im_\alpha(\eta_F)$ is the Fermi-Dirac integral of order $\alpha$ and $\text{n}$ is the carrier concentration. The matrix element for an impurity-electron scattering then becomes:

$$H_{k_x',k_x}^{m,n}(\vec{R}') = \int_R \frac{\|F_{n,k_x}(\vec{R})\|}{\sqrt{A}} \left( \frac{1}{L_x} \int_{-\infty}^{\infty} \frac{q_0^2 e^{iq_x x}}{4\pi\kappa_s\varepsilon_0} \frac{e^{-\sqrt{(\vec{R}-\vec{R}')^2+x^2}/L_D}}{\sqrt{(\vec{R}-\vec{R}')^2+x^2}} dx \right) \frac{\|F_{m,k_x'}(\vec{R})\|}{\sqrt{A}} d^2R$$

(27)

where the expression in the parenthesis is the Green's function of the infinite channel device. For a cylindrical channel, this becomes the modified Bessel function of second kind of order zero, $K_0(q, \vec{R}')$ [23, 67, 68, 69]. As in the case of phonon scattering, we use the probability density on each atomic site instead of the actual wavefunctions.

The matrix element in Eq. (27) is due to scattering from a single ionized impurity at $\vec{R}'$. The transition rate due to impurity scattering is computed after taking the square of the matrix element, multiplying by $N_I L_x$, the number of impurities in the normalized cross sectional area of the NW in the length of the unit cell, and integrating over the distribution of impurities in the cross sectional area of the NW (over $\vec{R}'$). The coordinates of the impurities are assumed to be the same as the coordinates of the atoms in the unit cell of the NW. A uniform distribution of dopant impurities over all NW atom positions is also assumed. The transition rate is then given by:

$$S_{n,m}^{imp.}(k_x, k_x') = \frac{2\pi}{\hbar} \left( \int_{R'} (N_I L_x) \left| H_{k_x',k_x}^{m,n}(\vec{R}') \right|^2 d^2R' \right) \delta(E_m(k_x') - E_n(k_x)) \quad (28)$$

### D. "Selection rules" and general scattering considerations:

Elastic and inelastic scattering processes are considered. We consider bulk phonons, and follow the *bulk* Si scattering "selection rules". As an example, we demonstrate the selection of the final scattering states of a particular state $k_x$ in subband $n$ for the $D$=3nm [110] n-type and p-type NWs. Their dispersions are shown in Fig. 2a and Fig. 2b respectively. The dispersion of the n-type NW consists of six valleys, originating from the six equivalent energy ellipsoids of Si conduction band. These are projected onto the 1D *k*-space, and in the [110] orientation, three two-fold degenerate valleys are formed as shown in Fig. 2a. From bulk Si scattering selection rules the elastic processes (elastic phonons, SRS, impurity scattering) are only intra-valley, whereas inelastic ones (inelastic phonons) are only inter-valley. Since these two-fold degenerate valleys originate from different conduction band ellipsoids, we allow scattering within these valleys as follows: i) For elastic scattering, in a specific valley (either Γ or off-Γ) transitions are allowed only from even to even and odd to odd subband numbers. ii) Inelastic inter-valley scattering transitions are allowed from even to odd and vice versa only within a specific valley (Γ or off-Γ). For inelastic transitions between the different valleys as shown in Fig. 2a, the final state of each scattering event is selected according to whether the process is of *f*- or *g*-type. All six relevant phonon modes in Si are included [47, 70]. Similar procedure is followed for the final state selection in n-type NWs in the other transport orientations. For p-type, the selection rules are simpler. We do not distinguish between the heavy-hole, light-hole and split-off bands, and include only ADP and ODP processes as shown in Fig. 2b. Inter-band and intra-band scattering are all included in this case.

We note that these scattering "selection rules" are approximations for consistency in the treatment of phonons as it is done for bulk Si, rather than strict rules. In reality, these "selection rules" rise automatically from the symmetry of the wavefunctions and scattering potential. In that case, other transitions, although weak, would also be possible. Since we only store the probability density and not the actual wavefunctions, the symmetry is lost and these "selection rules" are imposed "by hand" in the simulation. However, this approximation, as well as the rest of the approximations we make such as bulk phonons, or the simplified SRS and impurity scattering models will affect our results only quantitatively. The qualitative behavior in terms of geometry and orientation effects we describe originates mostly from electronic structure and will only be slightly affected by these approximations.

## III. Transport properties of n-type NWs

**A. The TD function for n-type NWs:**

To understand the influence of scattering in 1D channels and compare between different NW types, in Fig. 3 we plot the transport distribution functions (TD) $\Xi(E)$ defined in Eq. (2) for n-type NWs. As we show further on, this quantity provides a convenient way to understand the results, the variations between them and relate back to the electronic structures of the NWs. It may provide a clearer metric compared to the scattering rates, for example, that are not a simple function of energy, but also depend on the valley the electron resides into, the subband, its degeneracy, and the $k$-vector of the particular state.

Figure 3a shows the ADP phonon limited TD of n-type NWs with $D$=3nm in three crystallographic orientations [100], [110] and [111]. In all cases, the TD is linear in the low energy region where only the first subband contributes to the TD (as shown by the linear dashed asymptotes). The slope of the TD at the linear region depends on the effective mass of the subbands and their degeneracy ($\eta$). It results from the fact that:

$$\begin{aligned}\Xi_n(E) &\propto \eta v_n^2(E) \tau_n(E) \frac{g_{1D}^n(E)}{A} \\ &\propto \eta v_n^2(E) \frac{A}{g_{1D}^n(E)} \frac{g_{1D}^n(E)}{A} \\ &= \left(\frac{\eta}{m^*}\right) 2E\end{aligned} \quad (29)$$

Here we have assumed that $\tau_n(E)$ follows a simple relation (at least for elastic, isotropic, and intravalley processes) as: $\tau_n(E) \propto A / g_{1D}^n(E)$. We have also used the relations $g_{1D}^n(E) \propto \sqrt{m^*/E}$ and $v_n(E) \propto \sqrt{2E/m^*}$ (valid for 1D and under the parabolic band approximation), where $m^*$ is the effective mass of the subband. The degeneracy and effective masses used to match the initial slopes of the NW TDs are: i) $\eta_{[100]}=4$, $m_{[100]}^*=0.28m_0$, ii) $\eta_{[110]}=2$, $m_{[110]}^*=0.17m_0$, and iii) $\eta_{[111]}=6$, $m_{[111]}^*=0.42m_0$. These numbers are in close agreement with the actual effective mass values of the subbands as computed in Ref. [13] for such thin NWs. In Fig. 3 the TDs are plotted for a carrier concentration of $n_0=10^{19}/cm^3$. The Fermi levels $E_F$ are all shifted to the same energy reference at $E=0eV$. This shifts the TDs for the different NWs in energy, but not their shape which is independent of carrier concentration for ADP scattering. The difference in the effective masses and degeneracies of the NWs in the different transport orientations will shift the relative TDs with respect to $E_F$. The [110] NW has the lowest $m^*$ and the lowest degeneracy, whereas the [111] NW the highest $m^*$ and the highest degeneracy. At the same carrier concentration the conduction band edge of the [110] NW will be closer to $E_F$ and that of the [111] NW will be farther, as also shown in Fig. 3a. Because of this, as we will show further down, the conductivity of the $D=3nm$ [110] NW will be the highest at a given carrier concentration.

As the diameter increases, more and more subbands come into play. The TDs of the [111] n-type NWs of $D=3nm$ and $D=12nm$ are shown in Fig. 3b. Indeed, the TD of the thicker NW has many more peaks close by in energy. For the same carrier concentration $n_0=10^{19}/cm^3$, the TD of the thicker NW is located closer to $E_F$ than that of the narrow NW. This may appear counter intuitive since one would have expected that the smaller number of subbands in the thinner NW will force it to be closer to $E_F$. The placement of the subband edges with respect to $E_F$, however, depends on the number of states that need to be filled. For a specific diameter (at a specific carrier concentration), closely packed subbands will increase the distance $E_C$-$E_F$ of the conduction band from $E_F$. As the diameter is reduced, however, the number of subbands is reduced as well. If the number of subbands in the energy region of relevance decreases linearly with the cross section area of the NW, then $E_C$-$E_F$ will stay the same. Once a single subband remains, reducing the diameter does not further reduce the number of subbands. At the same 3D carrier concentration, $E_C$-$E_F$ increases (less carriers reside in the smaller area). The wider NWs will, therefore, have a higher conductivity than the thinner ones. The slopes of the TDs in the linear regions for narrow and wide NWs are the same, depending only on the effective mass and degeneracy of the subbands. The increase in the waveform overlaps with diameter reduction does not affect the TDs because its inverse dependence on the cross sectional area is canceled by the normalized DOS ($\frac{g_{1D}^n(E)}{A}$) in the definition of the TD in Eq. (2).

Once more scattering mechanisms are considered in the calculation of the TD, its magnitude gradually decreases as shown in Fig. 3c for the case of the n-type, [111], $D$=3nm NW. Here we include ADP, and sequentially add IVS, SRS and impurity scattering, assuming a rather large impurity concentration of $n_0=10^{19}$/cm$^3$. For this very thin $D$=3nm NW, SRS is a very strong scattering mechanism, reducing the TD by more than a factor of ~3X from to the phonon-limited value. Impurity scattering is the dominant scattering mechanism, similar to the bulk Si case for such high impurity concentrations [71, 72]. The inset of Fig. 3c shows the mean free path (MFP) for electrons in the first subband of the $D$=3nm, [111] NW. The MFP is defined as $\lambda_n = 2\pi\hbar\Xi(E)/M$, where $M$ is the number of modes [73]. The MFP peaks around ~50nm for the phonon-limited conditions, but drops to ~20nm as SRS is included and further on to ~5nm when impurity scattering is additionally included. Persson *et* al. in Ref. [16], and Lherbier *et* al. in Ref. [74] using the Landauer transport formalism and the Kubo-formula, report that the SRS-limited electron MFPs for the first subband of NWs of small diameters peak around ~40-70nm depending on the orientation. These numbers are in quantitative agreement with our calculations for the SRS-limited MFP peaks. For impurity scattering limited MFPs, Markussen *et* al. [75] report for narrow n-type [100] NWs doped at $n_0=10^{19}$/cm$^3$ MFPs that vary from ~10nm at low energies to ~50nm at higher energies. These values are also quantitatively close to our impurity scattering limited results at the same impurity concentration for the [100] NW (not shown here). We note however, that although the *peaks* of the MFPs in those works seem to agree with our results, those works included the effect of localization which is stronger at lower energies, and their low energy carrier MFPs had lower amplitudes.

B. **Conductivity and low–field mobility for n-type NWs:**

From the TD, the electrical conductivity and mobility can be extracted. Figure 4a and 4b show the conductivity and mobility, respectively, for the n-type, [100] NW of $D$=3nm as a function of the electron concentration for various combinations of scattering mechanisms. Phonon-limited (solid), phonon and SRS (dashed), phonon and impurity scattering (dashed-dotted), and phonon, SRS and impurity scattering (solid-red) results are shown. SRS is indeed a strong mechanism for this narrow NW, and reduces its conductivity and mobility by almost 4X compared to the phonon limited values. For impurities, we assume that the concentration is equal to the carrier concentration in the channel. Similar to bulk Si [71, 72], impurity scattering becomes important at concentrations above $n_0$=$10^{16}$/cm$^3$. At concentrations of $n_0$=$10^{20}$/cm$^3$ it can reduce the conductivity and mobility by an order of magnitude from the phonon-limited values.

The maximum phonon-limited mobility reached in Fig. 4 is $\mu_n$~400cm$^2$/Vs, much lower than the bulk value of $\mu_n$=1400cm$^2$/Vs. Clearly, the mobility suffers at this reduced diameter. To illustrate the effect of diameter on the mobility, Figure 5 shows the low-field mobility for the NWs in the [100] (square-blue), [110] (triangle-red) and [111] (circle-green) transport orientations as a function of the NW diameter ranging from $D$=12nm down to $D$=3nm. Figure 5a shows the phonon-limited mobility at low carrier concentrations (taken from the left-most constant mobility regions of Fig. 4b). The mobilities for larger diameters are below the bulk electron mobility $\mu_n$=1400cm$^2$/Vs, since the phonon coupling constants used are larger than the usual bulk values. Using the lower value will increase the large diameter mobilities and bring them closer to $\mu_n$=1400cm$^2$/Vs. A clear decrease in mobility is observed as the diameter is reduced. This is explained by the fact that at the same carrier concentration the TD of the smaller diameter NW is farther from $E_F$ as shown in Fig. 3b. (Although the TD in Fig. 3b is plotted at high concentration, the same behavior is observed for lower concentrations as well). We note here that the wave form factor overlap increases as the diameter is reduced since it is inversely proportional to the NWs' area ($1/A_{nm} \sim 1/Area$), and thus enhances phonon scattering. Its influence, however, is more prominent at diameters below $D=3$nm, where only a single subband contributes to transport. Assuming a single-degenerate valley, the mobility is roughly proportional to:

$$\mu = \frac{\sigma}{q_0 n} \propto \frac{\int_E v_n^2(E) \frac{A}{Mg_{1D}^n(E)} \frac{Mg_{1D}^n(E)}{A} \left(-\frac{\partial f(E)}{\partial E}\right) dE}{M \int_E \frac{g_{1D}^n(E)}{A} f(E) dE}$$

$$= \left(\frac{A}{M}\right) \frac{\int_E v_n^2(E) \left(-\frac{\partial f(E)}{\partial E}\right) dE}{\int_E g_{1D}^n(E) f(E) dE}$$

(30)

where $M$ is the number of subbands. As the area decreases, the number of subbands also decreases usually linearly for the thicker NWs. The overlaps increase, but the number of subbands decrease linearly as well, leaving less final states for scattering available, decreasing the scattering rates and leaving the mobility unaffected. At some point only a single subband participates in transport ($M=1$). Usually for Si at room temperature this happens at $D=3$nm. Further reduction of the diameter will scale the mobility linearly. Again using $g_{1D}^n(E) \propto \sqrt{m^*/\tilde{E}}$ and $v_n(E) \propto \sqrt{\tilde{E}/m^*}$, where $\tilde{E} = E - E_0$, with $E_0$ being the energy of the subband edge, then:

$$\mu \propto \left(\frac{A}{M}\right) \frac{\int_{E_0}^{\infty} \tilde{E}/m_{eff} \left(-\frac{\partial f(E)}{\partial E}\right) dE}{\int_{E_0}^{\infty} \sqrt{\frac{m_{eff}}{\tilde{E}}} f(E) dE}$$

$$= \frac{1}{m_{eff}^{3/2}} \left(\frac{A}{M}\right) \tilde{F}(\eta_F)$$

(31)

where $\tilde{F}(\eta_F)$ is some function of $\eta_F = E_F - E_0$, where $E_F$ is the Fermi level. In the non-degenerate limit, this function is independent of $\eta_F$ and bandstructure. The mobility, therefore, has an effective mass dependence $m_{eff}^{-3/2}$ as also mentioned in Ref. [19] and consequently, a third power dependence on the injection velocity ($v_{inj.}$):

$$\mu \propto m_{eff}^{-3/2}\left(\frac{A}{M}\right) \propto v_{inj.}^{3}\left(\frac{A}{M}\right). \tag{32}$$

Some degree of orientation dependence is observed for both, narrow and wide NWs. The electron mobility of the [100] and [111] NWs in Fig. 5(b) drops by ~3x and ~4x, respectively, whereas that of the [110] NW is less affected, and drops by only ~1.2X. The effective mass of the lowest subband of the [110] NW decreases with quantization from $m^*=0.19m_0$ to $m^*=0.16m_0$ [13]. This increases the carrier velocities and reduces the scattering rates. From the TD point of view, i) it increases its slope, and ii) it tends to shift the band edges closer to $E_F$ at a particular carrier concentration. This slightly compensates the fact that the band edge shifts farther from $E_F$ with diameter reduction. On the other hand, the mass of the lowest bands of the [100] NW increases from $m^*=0.19m_0$ to $m^*=0.27m_0$ [13], which further pushes the band edges away from the Fermi level. Comparing the mobility of the smaller NWs with $D$=3nm, the one of the [110] NW is the largest, ~2X higher than the other NWs, followed by the [100] NW, whereas that of the [111] is the lowest. The ordering is the same as that of the TD in Fig. 3a. At $D$=12nm, the [100] NWs have the largest mobility, ~60% higher than the other orientations, followed by the [111] and finally the [110] NWs. The [110] NW has the lowest mobility because it's heavier four-fold degenerate off-Γ valleys with $m^*=0.55m_0$ [13] are more heavily quantized, and as the diameter increases they lower close to the Γ valley faster. This makes the [110] NW suffer more from inter-valley scattering, as well as makes the dispersion overall "heavier", and the mobility is reduced compared to the other NWs. This is a similar effect as the behavior of mobility of bulk GaAs at high field.

In Fig. 5b SRS is additionally included. The phonon-limited mobility of Fig. 5a is also shown in fainted lines for comparison. SRS degrades the mobility, especially for the smaller diameters by ~3X, similarly to what is shown in Fig. 4b. This large drop for the smaller diameters is explained from the large shift $\frac{\Delta E_C}{\Delta D}$ in the subband edges as the as the diameter is scaled (see Fig. 1).

In Fig. 5c impurity scattering is additionally included. In this case a rather large impurity concentration of $n_0=10^{19}$/cm$^3$ is considered. The phonon-limited results at carrier concentrations of $n=10^{19}$/cm$^3$ are included in fainted lines for comparison purposes. As also shown in Fig. 4, at such a high concentration, impurities are the dominant scatterers and reduce the mobility by an order of magnitude compared to the phonon-limited mobility. At larger diameters, the [110] NW is affected the least compared to NWs of other orientations (but still severely). A possible reason is that the overall "heavier" dispersion subbands of this NW are less affected by impurity scattering. The impurity scattering matrix element for bulk impurities (or large NW diameters in this case) is proportional to $\left(1+q^2 L_D^2\right)^{-1}$ [47]. Therefore, the momentum relaxation time includes terms proportional to $\tau \propto q^2, q^4$. The larger the momentum transfer, i.e. the heavier the band effective mass, the larger $\tau$ and TD are, and the mobility suffers less.

Should be noted that electron-impurity interaction can in reality be much stronger in NWs than in bulk channels. The linearized BTE applied here does not capture any of the quantum coherence effects detrimental to the conductivity, such as carrier localization [16, 75], or effects as backscattering Fano resonances due to potential wells formed around ionized dopants [76, 77], especially when the impurities are placed near the center of the channel [78, 79]. In addition, in describing impurity scattering we have used the expression for bulk dielectric screening, which might not fully apply for the 1D geometry [23]. These factors can potentially make the impurity scattering stronger and degrade the mobility even further depending on the structure and operating condition details, especially for the thinner NWs. Localization effects are strongly weakened when decoherence is strong (i.e. phonon scattering in room or higher temperatures, or electron-electron interactions at high carrier concentrations [80]) and when the phase relaxation length is short, shorter than the distance between successive scatterers [80]. For the thinnest $D=3$nm NWs, however, under strong SRS and high impurity concentrations $n_0=10^{19}$/cm$^3$ (this results in one impurity per 15nm), the MFPs can be quite short (insets of Fig. 3c and Fig. 6c). The SRS and impurity scattering MFPs can actually be shorted than the MFPs of the phase relaxing phonon-limited mechanisms. Therefore, coherent scattering events might be of importance and can limit the conductivity even further under such conditions. At such high carrier concentrations, on the other hand, electron-electron interactions can become important, and the phase relaxation length can strongly decrease.

We consider here the technologically relevant applications of NWs in electronic devices where the regime of localization has to be avoided. Therefore, we focus on the diffusive transport regime, rather than the localization regime and do not discuss the exact borders between the two regimes. Calculations of the effect of localization on the mobility of narrow NWs [16, 74, 75], however, show similar orientation trends and qualitative agreement to our results both for electrons and holes. We, therefore, believe that even if localization is present, our basic conclusions about geometry and orientation anisotropy will not be altered (although the absolute mobility values might decrease).

## IV. Transport properties of p-type NWs

In this section, we perform the analysis of transport properties in p-type NWs. As we will show, in this case the results show much stronger anisotropy. More importantly, the trends in some cases are opposite from what is observed for n-type NWs, i.e. diameter scaling can actually provide enhancements in the conductivity and mobility.

C. **The TD function for p-type NWs:**

As earlier in Fig. 3 for n-type NWs, Fig. 6 shows the TD for the p-type NWs. Figure 6a shows the ADP phonon limited TD of NWs in the three crystallographic orientations [100], [110] and [111] for diameter $D$=3nm. Again, in all cases, the TD is linear in the low energy region where only the first subband contributes to the TD. The slope of the TD depends on the effective masses of the subbands and their degeneracy. At slightly higher energies non-parabolicity effects appear and the TD deviates from the linear asymptotes with slope ~ $\eta/m^*$ ($\eta$=1 in the p-type cases). The effective masses used to plot the linear asymptotes are: i) $m_{[100]}^*=0.9m_0$, ii) $m_{[110]}^*=0.22m_0$, and iii) $m_{[111]}^*=0.17m_0$. These numbers agree very well with estimated averaged effective mass values for these NWs as computed in Ref. [46], where we first reported on the p-type mobility in NWs. It is obvious from here, that the [111] NWs and secondly the [110] NWs have a large advantage compared to the [100] NWs. At the same carrier concentration of $p_0=10^{19}/cm^3$ not only their TDs are closer to the Fermi level, but they are also much higher. As we will show below, the conductivity and mobility of the $D$=3nm [111] and [110] NWs will be much higher than that of the [100] NW.

As the diameter increases more subbands come into play. Figure 6b shows the TD of the [111] *p-type* NW for $D$=3nm and $D$=12nm. The TD for the $D$=12nm NW has many more peaks close by in energy and is much lower than the one for $D$=3nm. The TD of both the $D$=3nm and $D$=12nm NWs, however, have approximately the same separation from $E_F$. The narrower NWs will therefore have a higher mobility than the thicker ones as we will show below, in contrast to what is observed for n-type NWs.

Once more scattering mechanisms are considered in the calculation of the TD, its magnitude gradually decreases as shown in Fig. 6c for the case of the [111] p-type $D$=3nm NW. Here we include ADP, then add ODP and finally add SRS. Although the strengths of the ODP scattering and SRS depend on the values used for the deformation potential and the $\Delta_{rms}$ value, respectively, these two scattering mechanisms turn out to be strong, reducing the TD by more than a factor of ~4X compared to the ADP phonon limited case, similar to the case of n-type NWs in Fig. 3. The inset of Fig. 6c shows the mean free path for the holes in the first subband of the NW, derived in the same way as above in Fig. 3c for electrons [73]. The MFP peaks around ~100nm for the phonon-limited and SRS limited conditions, higher than in the case of electrons. This is in qualitative agreement with the results of Ref. 16 for SRS limited conditions for the [111] NWs.

### D. Conductivity and low–field mobility for p-type NWs:

The TD features determine the electrical conductivity of the NWs shown in Fig. 7. Here the conductivity of the p-type NWs with $D$=3nm (solid) and $D$=12nm (dashed) in the three crystallographic orientations [100] (blue), [100] (red), and [100] (green) as a function of the hole concentration is shown. Figure 7a demonstrates the phonon-limited conductivity. For p-type [111] and [110] NWs the conductivity is much higher for $D$=3nm than for $D$=12nm. As explained in Ref. [46], this advantage for the thinner NWs originates from a large curvature increase in the subbands of the [111] and [110] NWs with confinement. This is shown in the dispersions of the [111] NW in the Inset of Fig 7b for the $D$=12nm (left) and $D$=3nm (right) sides. Thus, the effective masses are reduced with scaling, which increases the hole velocities. This reduction in the effective mass is responsible for the higher TD of the thinner NW in Fig. 6b, but also the fact that it does not shift further from the $E_F$ compared to the thicker NW. At carrier concentrations around $n$=5x10$^{19}$/cm$^3$ the conductivities reach a maximum and then start to decrease as more and more of the slower carriers from smaller curvature subband regions start to contribute to transport (lower energy subband regions in the Inset of Fig 7b-right side). The conductivity of the [100] p-type NW stays low, and similarly to n-type NWs, it is reduced at smaller diameters. Figure 7b shows the conductivity when SRS is included in addition to phonon scattering. The improvement in the [110] and [111] cases is large enough to compensate for the detrimental effect of SRS, and the conductivity still benefits from diameter scaling.

Figure 8 shows the low-field mobility as a function of the NW diameter from $D$=12nm down to $D$=3nm for the [100] (square-blue), [110] (triangle-red) and [111] (circle-green) transport orientations. Figure 8a shows the phonon-limited mobility. The mobilities for larger diameters are below the bulk hole mobility of $\mu_p$=450cm$^2$/Vs, since the optical phonon scattering amplitude used is larger than the one used to match the bulk mobility. Using the lower value will increase the mobilities at large diameter and bring them closer to $\mu_p$=450cm$^2$/Vs [46]. Nevertheless, our results demonstrate a clear trend of increasing mobility as the diameter decreases. The increase is observed in the cases of [110] and mostly in the case of the [111] NWs, where the phonon-limited mobilities reach up to ~1300 cm$^2$/Vs (in agreement with the work of Refs [19, 30]) and ~2200

cm$^2$/Vs, respectively. These are a ~3X and ~5X increase from the bulk mobility values, respectively. Comparing the values at $D$=3nm and $D$=12nm, the increase is about 8X. As we have shown in Ref. [29], the carrier velocity in p-type [110] and [111] NWs can increase by a factor of ~2X as the diameter is reduced down to $D$=3nm. From Eq. (32), the mobility is proportional to $v_{inj.}^3$, which justifies the ~8X increase in the mobility. It is a demonstration of how the length scale degree of freedom can provide a mechanism of performance enhancement in nanostructures. Under strong scaling at $D$=3nm, the effective masses of these NWs can be even lower than the electron transverse effective mass in the conduction band ($m_t^*$ = 0.19m$_0$) [16, 30, 46]. This explains why their mobilities are much higher than that of bulk *p-type* channels (at least the phonon-limited mobility, since $\mu = q\tau/m^*$). In such case, the mobility benefits from: i) the mass decrease, and ii) an increase in $\tau$, as the lower density of states of the lighter subbands reduces the scattering rates. On the other hand, the mobility of the [100] NWs is by far the lowest for all diameters, and additionally it decreases with diameter scaling.

When SRS is included, in Fig. 8b, at larger diameters the two curves coincide since the band edges do not significantly shift in channels with thicknesses larger than 6nm (Fig. 1). For the smallest diameters, however, SRS degrades the mobility by almost ~2X in all NW cases. The magnitude of this degradation is in qualitative agreement with other works on the mobility degradation of SRS in nanodevices [23], although it strongly depends on the parameters used in Eq. (24). Nevertheless, the [110] and [111] channels, that provide increasing carrier velocities as the channel width decreases, can partially compensate for SRS and still provide mobility benefits. As an example, we mention that an effective way to design high efficiency nanostructured thermoelectric devices is to scale the feature sizes in order to reduce the lattice thermal conductivity. In that scope, efficient thermoelectric devices based on silicon NWs have already been demonstrated [11, 12]. High electronic conductivity though is still needed. The [110] and [111] *p-type* NE channels, in which the mobility and conductivity increase with feature size reduction, might be suitable for such applications. On the other hand, for the [100] NWs, the mobility is always below ~100 cm$^2$/Vs as shown for the [100] mobility data in the Inset of Fig. 8b. Similar conclusions regarding anisotropy have been reported by Person *et* al.

in Ref. [16] using a different transport formalism. In that work, using the Landauer transport formalism the authors showed that the mean free path of holes in SRS dominated p-type [111] NWs of $D$=2nm peaks at ~70nm, it is higher than that of the holes in [110] NW (~20nm), and significantly higher than that of holes in NWs in the [100] orientation.

In Fig. 8c the mobility is computed including phonons, SRS and additionally impurity scattering. A large impurity concentration of $p_0=10^{19}/cm^3$ is included. The mobility indeed drops significantly by almost 10X compared to the phonon-limited mobility, but the diameter trend and the orientation dependence is retained, namely that the [111] and [110] NWs still benefit from diameter scaling.

# V. Discussion

Similar transport studies are also discussed in the literature for either NWs of smaller diameters, fewer scattering processes [16, 30, 31], or different channel structures and materials [81, 82]. In this work, we include all basic scattering mechanisms, and much larger diameters up to $D$=12nm. To be able to couple TB electronic structures and Boltzmann transport calculations for such large structures, several approximations are made: i) The assumption of deformation potential theory and dispersionless bulk phonons instead of confined phonons, ii) the probability density instead of the actual wavefunctions was used, iii) a simplified treatment of SRS and impurity scattering, iv) unrelaxed unstrained structures instead of reconstructed ones were considered. These are reasonable approximations and only affect our results quantitatively. The quantitative behavior in terms of geometry and orientation effects we describe originates mostly from electronic structure and is only weakly affected by these approximations.

### A. **The possible effect of surface relaxation and passivation:**

Approximation (iv) might be the one with the biggest qualitative influence on our results. Our conclusions are electronic structure based, and relaxation might have some effect on the electronic structure especially for the narrower $D$=3nm NWs. It is useful, therefore, to provide an estimate of how it can affect our results for the narrowest NWs.

Several theoretical reports in the literature discuss the effect of passivation and relaxation of NW surfaces using *ab-initio* methods [83, 84, 85, 86]. The general conclusion is that passivation does not modify the physical structure in any noticeable way even for ultra-thin NWs down to 1nm in diameter [84]. The Si-Si surface bond lengths change only by <1.5%, whereas in the cores of the NWs the bond variation is negligible (<0.1%) [83, 87]. The relaxation along the NW axis for narrow $D$=3nm NWs in various orientations is also overall less than 1.3% [85]. The bandgaps, on the other hand, change upon passivation, and they are strongly influenced by the choice of passivation agents (some of the more common ones are -H, -OH, and –NH$_2$, and SiO$_2$) because of the different surface charge centers introduced [86].

Regarding the effective masses of the subbands, which have the main influence on the electronic properties we investigate, Vo *et al.* [83] have calculated the effect of hydrogen passivation and reconstruction on the electron and hole masses in Si NWs of diameters 1-3nm. For the 3nm NWs, they show that the effective masses of the different oriented NWs change upon relaxation as follows: For electrons: [100] ~15% increase, [110] ~7% decrease, [111] ~15% increase. For holes: [100] ~30% decrease, [110] no change, [111] no change. The p-type [100] NWs are affected the most. From Eq. (31) a 30% reduction in the effective mass will roughly result in $\frac{\Delta \mu_n}{\mu_n} \propto \left( \frac{\Delta m_n^*}{m_n^*} \right)^{-3/2}$ ~70% increase in the mobility. The mobility we report for this NW is much lower than that of the other NWs, anyway, and therefore our anisotropy conclusions will not be altered. In the case of electrons in [100] and [111] NWs, the NW mobility might be ~30% lower than what our calculations show once relaxation is considered. Again, for this case, our mobility results for these *n*-type $D$=3nm NWs are lower than the [110] NW, and therefore the conclusion with regards to anisotropy are still valid.

Regarding the transport properties, in another work Vo *et al.* [31] reported conductivity calculations for $D$=1.1nm NWs in different transport orientations. Although a different transport model and different diameters were used, the conductivity trends they present are very similar to the mobility trends we present in our work (but for the $D$=3nm NWs). Specifically for electrons in [110] NWs, Liang *et al.* [88] investigated the effect of reconstruction in $D$=1nm NWs, and found that its effect on the ballistic performance was minimal because the variation of the effective mass was negligible upon reconstruction, as also verified by Vo *et* al in Ref. [83].

It is therefore believed that reconstruction might somewhat affect the magnitude of the mobility we calculate in some cases, and noticeably only for the $D$=3nm p-type [100] NW. The effect, however, does not seem to be strong enough to alter the anisotropy behavior we describe.

**B. <u>1D to 3D transition:</u>**

By including all basic scattering mechanisms and considering diameters up to $D$=12nm, we can have an indication of the possible length scale at which the transition from 1D like behavior ($D$=3nm) to 3D bulk-like appears. To identify this we need to look at the diameter scale where critical electronic parameters saturate, or saturate close to their bulk value. From the results presented in Fig. 1 for the band edge shifts and Fig. 5 and Fig. 8 for the mobility versus diameter, we can observe that the larger changes appear for diameters below $D$~8nm, both for the conduction and valence bands. Above $D$~8nm, these quantities begin to saturate asymptotically. From our previous work in Ref. [29] we show that the bandstructure carrier velocities as well begin to saturate at diameters above $D$~8nm, or slightly larger than that for the cases of [110] and [111] holes [29, 46]. Very similar diameter behavior is also observed for the effective masses of the bands (or the band curvature in the case of p-type NWs) [13, 46]. It therefore seems that in order to observe low-dimensional bandstructure effects the diameter needs to be scaled roughly below $D$~10nm.

On the other hand, the bulk Si mobility is isotropic due to symmetry reasons. Therefore, one should expect that in order to clearly observe a transition from 1D to 3D, the mobility for all NW orientations needs to converge to the same value. At $D$=12nm the mobilities of the different oriented NWs still do not converge to a single value (Fig. 5, Fig. 8), meaning that they still feel the influence of the anisotropy in the carrier bandstructure velocity as shown in Ref. [29]. The reason is that still large enough band separations exist as shown in Ref. [89]. At $D$~12nm the subbands are not yet all close together to form solid 3D valleys. This effect does not show up in the masses, carrier velocities, or bands energies, but it shows up in the *DOS(E)* and mobility.

We can therefore conclude that although as the diameter is increased to $D$~10nm a large degree of 3D features appear in the electronic structure, the clear transition to 3D will gradually happen at larger diameters, presumably even beyond $D$~30nm.

# VI. Conclusion

A method to efficiently couple atomistic electronic structures from TB calculations with the linearized Boltzmann formalism for calculation of transport properties of Si NWs is described. The 20 orbital $sp^3d^5s^*$-spin-orbit–coupled TB model was utilized. Phonon, SRS and impurity scattering are included. The full *k*-vector, subband and energy dependence of the momentum relaxation rates and of the carrier velocities are included in the calculations. Within reasonable approximations, the method is computationally robust compared to fully *ab-initio* methods, and although our simulations can include up to ~5500 atoms (for NWs up to 12nm in diameter), they can be performed on single, or limited number of CPUs.

Using this approach, we present a comprehensive investigation of the low-field mobility in silicon NWs considering: i) n- and p-type NWs, ii) [100], [110], and [111] transport orientations, and iii) diameters from $D$=12nm down to $D$=3nm. We find that phonon and SRS become stronger as the diameter is reduced and drastically degrade the mobility by up to an order of magnitude depending on the orientation. This is observed for the [100] p-type NWs and all n-type NWs with the [110] ones less affected. Impurity scattering as in the case of bulk Si, seems to be a strong mechanism at high concentrations.

In the cases of p-type [110] and [111] NWs, on the other hand, confinement significantly increases the subbands' curvature and hole velocities. As a consequence, the phonon-limited mobility in these channels largely increases by a factor of ~8X as the diameter scales from $D$=12nm down to $D$=3nm. This overcompensates the detrimental effect of enhanced phonon scattering and SRS and the mobilities of the [110] and [111] NWs benefit from cross-section scaling. Our results agree with other computational studies [19, 30], and could potentially provide explanations for the large mobility enhancement in NWs that has been observed experimentally [38], as well as insight into the strong mobility orientation effects in ultra-scaled MOSFET device measurements [22].

## Acknowledgements

This work was supported by the Austrian Climate and Energy Fund, contract No. 825467.

Figure 1:

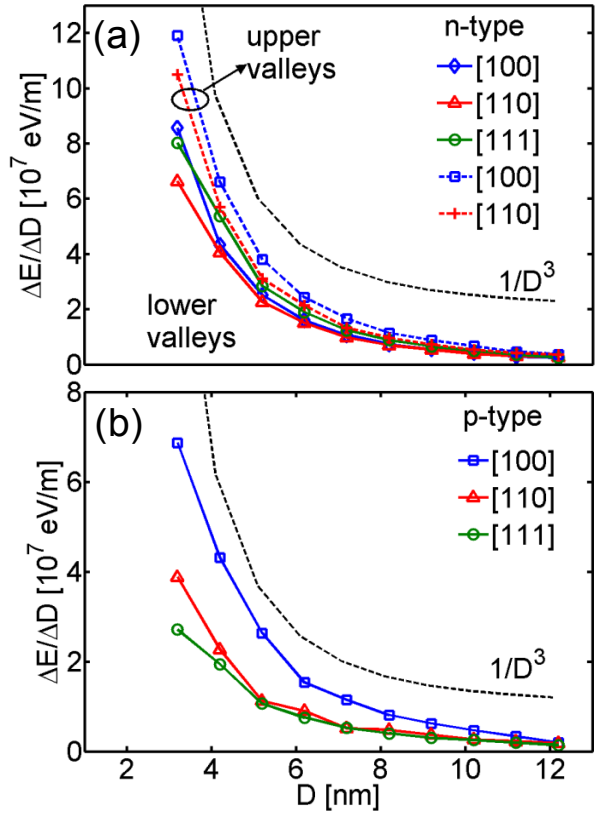

Figure 1 caption:

The change in the band edges as a function of diameter. Results for NWs in the [100] (square-blue), [110] (triangle-red), and [111] (circle-green) transport orientations are shown. This variation defines the SRS strength. (a) Conduction band. Results for the lower valley (solid), and upper valley (dash) are shown. (b) Valence band. The dashed-black line in (a) and (b) indicates the $D^{-3}$ law.

Figure 2:

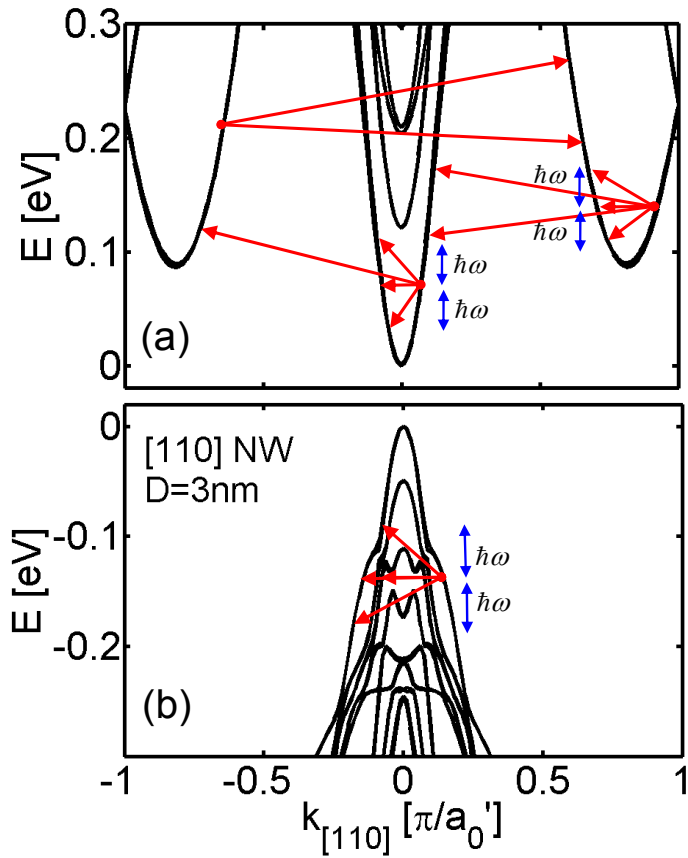

Figure 2 caption:

Dispersions of the $D$=3nm, [110] NW with the scattering mechanisms indicated. (a) n-type NW. Intra-valley elastic and inter-valley inelastic (IVS) processes are considered (between the three valleys), following the bulk silicon scattering selection rules. Elastic processes: ADP, SRS and impurity scattering. Inelastic processes: Three for each *f*- and *g*- phonon scattering processes. Each valley in the dispersion is double-degenerate. Following the bulk scattering selection rules, however, each of the valleys is considered independently. (b) p-type NW. Elastic and inelastic processes are considered within the entire bandstructure. Intra- and inter-valley scattering is considered.

Figure 3:

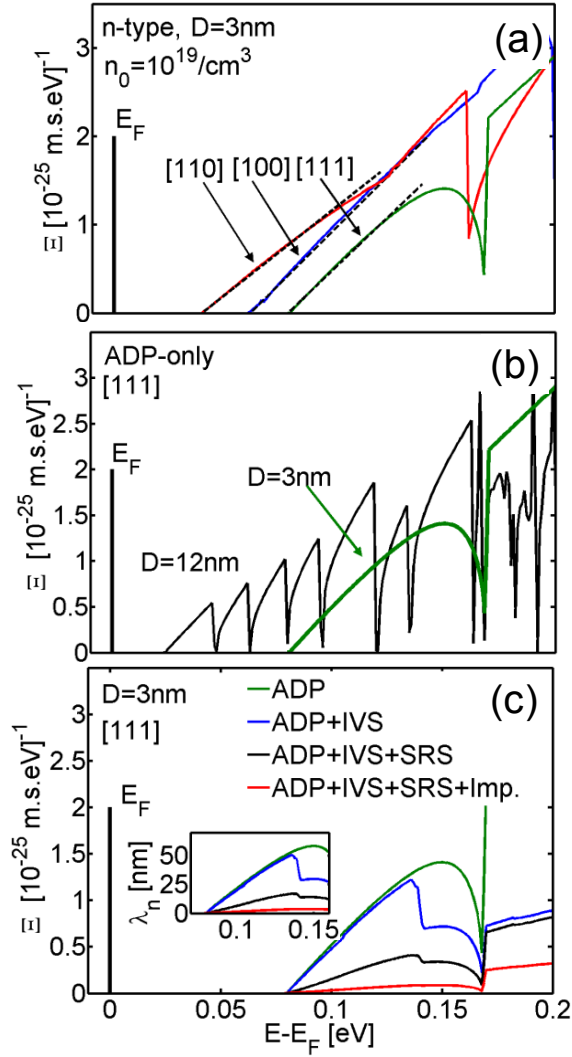

Figure 3 caption:

The transport distribution function (TD) $\Xi(E)$ for n-type NWs for carrier concentration $n=10^{19}/cm^3$. (a) ADP phonon-limited TD for the $D$=3nm NWs in the [100] (blue), [110] (red) and [111] (green) transport orientations. (b) ADP phonon-limited TD for the $D$=3nm (black) and $D$=12nm (green) [111] NWs. (c) TD for the [111], $D$=3nm NW under i) ADP phonon-limited, ii) ADP-IVS phonon-limited iii) ADP-IVS-SRS, and iv) ADP-IVS-SRS-Impurity scattering conditions. Inset of (c): The mean free path (MFP) for the electrons under the different scattering mechanisms.

Figure 4:

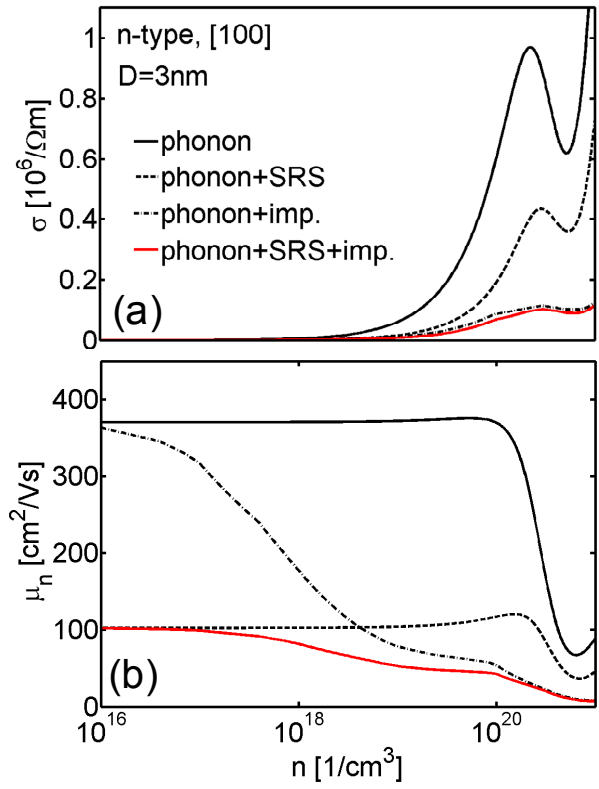

Figure 4 caption:

The electrical conductivity (a) and electron mobility (b) for the n-type [100], $D$=3nm NW versus the electron concentration. Phonon-limited (solid), Phonon-SRS (dashed), Phonon-Impurity (dash-dotted), and Phonon-SRS-Impurity (solid-red) scattering results are shown.

Figure 5:

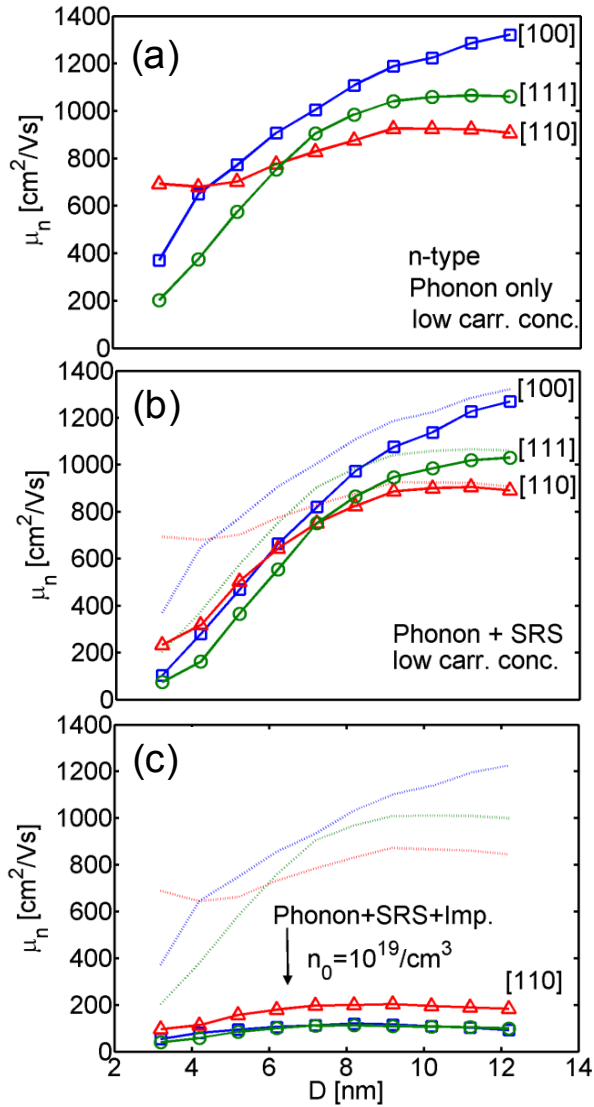

Figure 5 caption:

The low-field mobility for n-type NWs in [100] (square-blue), [110] (triangle-red) and [111] (circle-green) transport orientations versus the NWs' diameter. (a) Phonon-limited mobility. (b) Phonons and SRS are included. (c) Phonons, SRS and impurity scattering are included for impurity concentration $n_0=10^{19}/cm^3$. Dashed/fainted lines in (b) and (c): The phonon-limited results.

Figure 6:

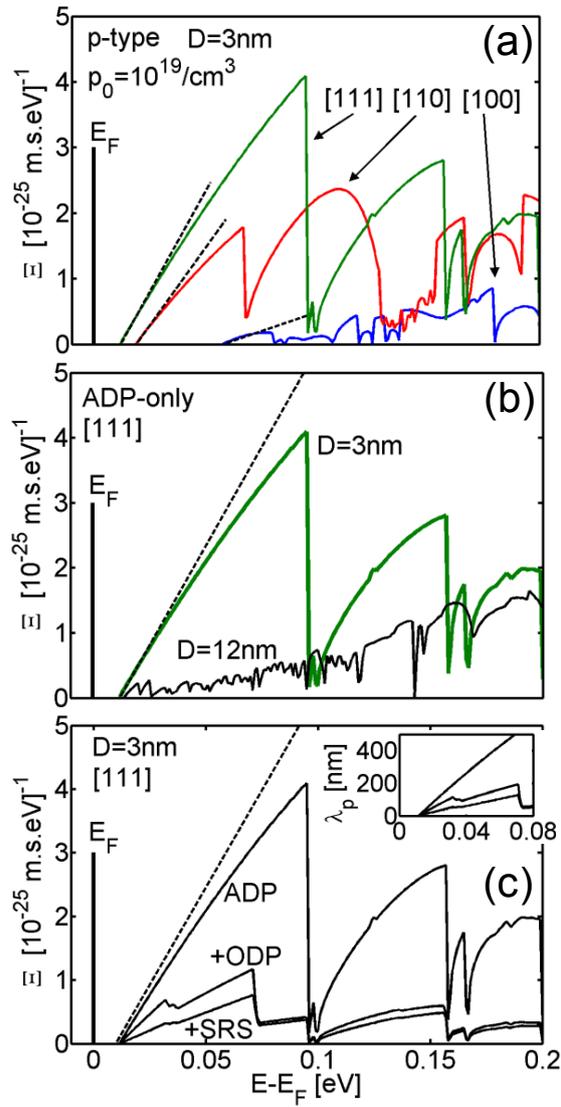

Figure 6 caption:

The transport distribution function (TD) $\Xi(E)$ for p-type NWs for carrier concentration $p=10^{19}/cm^3$. (a) ADP phonon-limited TD for the $D$=3nm NWs in the [100] (blue), [110] (red) and [111] (green) transport orientations. (b) ADP phonon-limited TD for the $D$=3nm (black) and $D$=12nm (green) [111] NWs. (c) TD for the $D$=3nm [111] NW under i) ADP phonon-limited, ii) ADP-ODP phonon-limited and iii) ADP-ODP-SRS scattering conditions. Inset of (c): The mean free path (MFP) for the holes under the different scattering mechanisms.

Figure 7:

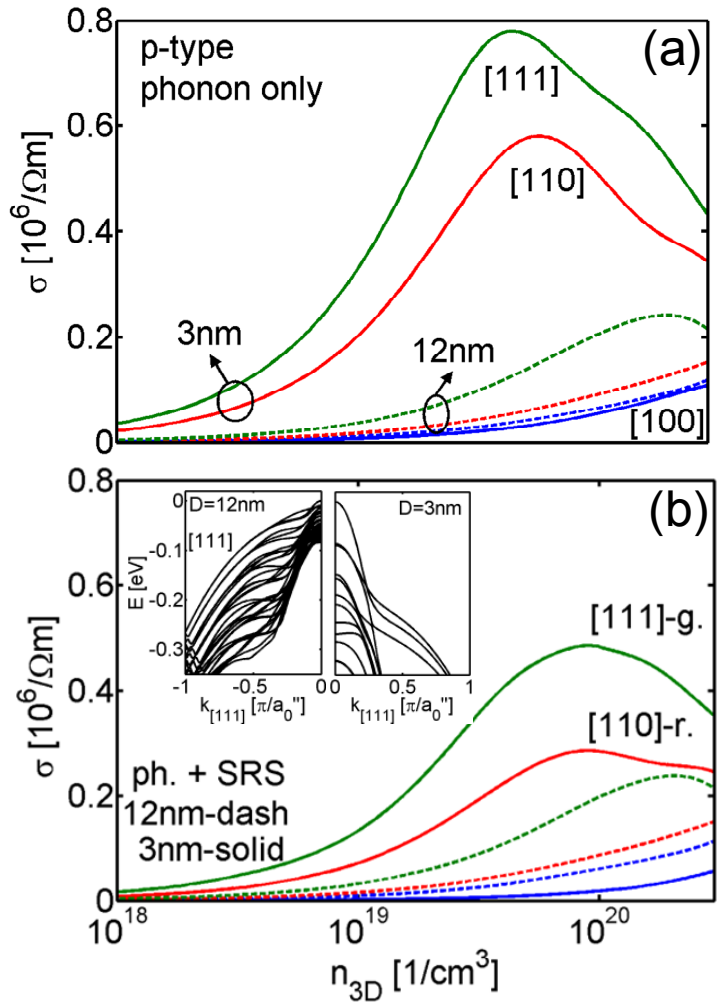

Figure 7 caption:

The electrical conductivity for p-type NWs in [100] (blue), [110] (red) and [111] (green) transport orientations for diameters $D$=3nm (solid) and $D$=12nm (dashed). (a) Phonon-limited mobility. (b) Phonons and SRS are included. Inset of (b): The dispersion of [111] NWs with, $D$ = 12 nm (left) and $D$ = 3 nm (right).

Figure 8:

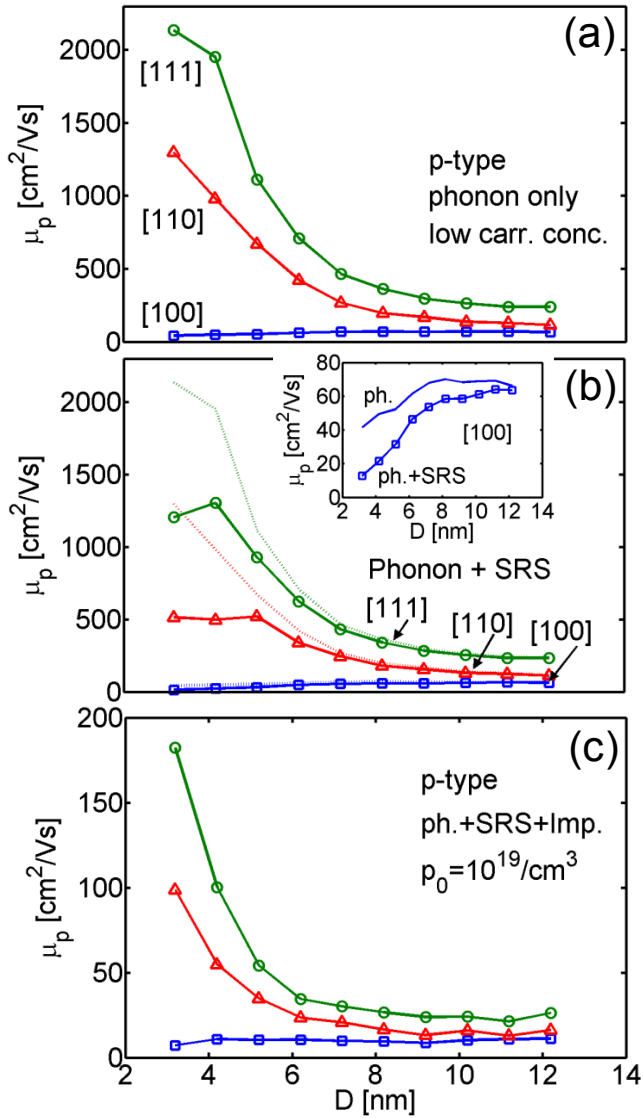

Figure 8 caption:

The low-field mobility for p-type NWs in [100] (square-blue), [110] (triangle-red) and [111] (circle-green) transport orientations versus the NWs' diameter. (a) Phonon-limited mobility. (b) Phonons and SRS are included. (c) Phonons, SRS and impurity scattering are included for $p_0=10^{19}/cm^3$. Dashed/fainted lines in (b): The phonon-limited results. Inset of (b): A zoom in the [100] NW results. The data in (a) and (b) are the same as in Ref. [46].